\documentstyle[epsfig,preprint,aps]{revtex}
\begin{document}
\draft
\title{Naked singularities in three-dimensions.} 
\author{G. Oliveira-Neto\thanks{Email:
gilneto@fisica.ufjf.br}} 
\address{Departamento de F\'{\i}sica,
Instituto de Ciencias Exatas,
Universidade Federal de Juiz de Fora,
CEP 36036-330, Juiz de Fora,
Minas Gerais, Brazil.}
\date{\today}
\maketitle

\begin{abstract}
We study an analytical solution to the Einstein's equations in
$2+1$-dimensions, representing the self-similar collapse of a
circularly symmetric, minimally coupled, massless, scalar field.
Depending on the value of certain parameters, this solution
represents the formation of naked singularities. Since our 
solution is asymptotically flat, these naked singularities may
be relevant for the weak cosmic censorship conjecture in
$2+1$-dimensions.
\end{abstract}

Since the work of M. W. Choptuik on the gravitational collapse 
of a massless scalar field \cite{choptuik}, many physicists 
have focused their attentions on the issue of gravitational 
collapse. An important arena where one can study the 
gravitational collapse is general relativity in $2+1$-dimensions. 
The great appeal of this theory comes from the fact that it 
retains many of the properties of general relativity in 
$3+1$-dimensions, but the field equations are greatly simplified 
\cite{jackiw}.

Presently, several black hole solutions in $2+1$-dimensional
general relativity are known \cite{mann}. Including the
first one to be discovered, the so-called BTZ black hole
\cite{banados}. All of them have an important property in
common: the presence of a negative cosmological constant,
which makes them asymptotically anti-de Sitter. 

Indeed, in a recent work it was demonstrated that a 
three-dimensional solution to the Einstein's equations,
with a positive cosmological constant ($\Lambda$), such that
the stress-energy tensor satisfies the dominant energy
condition, contains no apparent horizons \cite{ida}. The same
result applies to the case $\Lambda = 0$ in the presence of
matter fields. Therefore, this result explains the necessity
of a negative cosmological constant in order to a black hole
to form, in three-dimensional general relativity.

Based on \cite{ida}, we can say that the gravitational collapse
of ordinary matter, without a negative cosmological constant,
will never form a black hole in $2+1$-dimensional general
relativity. On the other hand, one can not exclude the possible 
formation of naked singularities as the result of the 
gravitational collapse, without a cosmological constant.

In fact, it has already been shown that the collapse of a disk
of pressureless dust in $2+1$-dimensions, without a cosmological 
constant, has as
one of its possible end states a naked singularity \cite{kuchar}.
The singularity is space-like and it is a scalar polynomial 
singularity \cite{tipler}, in other words, scalar polynomials 
constructed from the Riemann tensor are unbounded there. For an 
observer far from the collapse region the space-time is flat and
conical. Since the naked singularity defines a region of non-zero
measure in the parameter space of solutions and is asymptotically
flat, it may be considered as a counter-example to the weak
cosmic censorship conjecture \cite{penrose} in $2+1$-dimensions.

In the present paper we would like to present a solution to the
Einstein's equation, without a cosmological constant, representing
the self-similar, circularly symmetric, collapse of a minimally
coupled, massless, scalar field, in $2+1$-dimensions. As we shall
see this solution, depending on the value of certain parameters, 
represents the formation of naked singularities as the result of 
the collapse process.

We shall start by writing down the ansatz for the space-time metric.
As we have mentioned before, we would like to consider the circularly
symmetric, self-similar, collapse of a massless scalar field in 
$2+1$-dimensions. Therefore, we shall write our metric ansatz as,

\begin{equation}
\label{1}
ds^2\, =\, -\, 2 e^{2\sigma(u,v)} du dv\, +\, r^2(u,v) d\theta^2
\, ,
\end{equation}
where $\sigma(u,v)$ and $r(u,v)$ are two arbitrary functions to be
determined by the field equations, $(u,v)$ is a pair of null
coordinates varying in the range $(-\infty,\infty)$, and $\theta$
is an angular coordinate taking values in the usual domain 
$[0,2\pi]$.

The scalar field $\Phi$ will be a function only of the two null
coordinates and the expression for its stress-energy tensor
$T_{\alpha\beta}$ is given by \cite{wheeler},

\begin{equation}
\label{2}
T_{\alpha\beta}\, =\, \Phi,_\alpha \Phi,_\beta\, -\,
{1\over 2} g_{\alpha\beta} \Phi,_\lambda \Phi^{,_\lambda}\, .
\end{equation}
where $,$ denotes partial differentiation.

Now, combining Eqs. (\ref{1}) and (\ref{2}) we may obtain the
Einstein's equations which in the units of Ref. \cite{wheeler}
and after re-scaling the scalar field, so that it absorbs the 
appropriate numerical factor, take the following form,

\begin{equation}
\label{3}
2\sigma,_u r,_u\, -\, r,_{uu}\, =\, r (\Phi,_u)^2\, ,
\end{equation}
\begin{equation}
\label{4}
2 \sigma,_v r,_v\, -\, r,_{vv}\, =\, r (\Phi,_v)^2\, ,
\end{equation}
\begin{equation}
\label{5}
2 r \sigma,_{uv}\, +\, r,_{uv}\, =\, -\, r (\Phi,_u
\Phi,_v)\, ,
\end {equation}
\begin{equation}
\label{6}
r,_{uv}\, =\, 0\, ,
\end{equation}
The equation of motion for the scalar field, in these 
coordinates, is

\begin{equation}
\label{7}
2 r \Phi,_{uv}\, +\, \Phi,_{v} r,_u\, +\, \Phi,_{u} r,_v\,
=\, 0\, .
\end{equation}

The above system of non-linear, second-order, coupled, partial
differential equations (\ref{3})-(\ref{7}) has an analytical
solution if we impose that it is continuously self-similar. More
precisely, the solution assumes the existence of an homothetic
Killing vector of the form,

\begin{equation}
\label{7,5}
\xi\, =\, u {\partial \over \partial u}\, +\, \alpha v {\partial
\over \partial v}\, ,
\end{equation}
where $\alpha$ is a non-negative, real number associated with the
type of kinematic self-similarity. Following Coley \cite{coley}, 
$\alpha = 1$ characterizes a self-similarity of the first kind and
for all other values of $\alpha$ we shall have a self-similarity 
of the second kind. Here, we shall restrict our attention to 
$0 < \alpha \leq 1$. We can express the solution in terms of the 
variable $z = (\alpha v)^{1/\alpha}/u$.

Under these conditions our solution will be given by \cite{wang},

\begin{equation}
\label{8}
r(u,v)\, =\, \beta (\alpha v)^{1/\alpha}\, +\, \gamma u\, ,
\end{equation}
\begin{equation}
\label{9}
\sigma (u,v)\, =\, \left({1 - \alpha\over 2}\right)\ln{\left({r
\over u}\right)}\, +\, \sigma_0\, ,
\end{equation}
and the scalar field has the following values,
\begin{equation}
\label{9,5}
\Phi (u,v)\, =\, 2 (\alpha - 1)^{1/2} \arctan  
\sqrt{{\beta(\alpha v)^{1/\alpha}\over \gamma u}}\, ,
\end{equation}
for $\gamma/\beta > 0$ and
\begin{equation}
\label{10}
\Phi (u,v)\, =\, (1 - \alpha)^{1/2} \ln \left[{ 
\sqrt{(\gamma/\beta)u}\, -\, \imath 
\sqrt{(\alpha v)^{1/\alpha}}\over 
\sqrt{(\gamma/\beta)u}\, +\, \imath 
\sqrt{(\alpha v)^{1/\alpha}}}\right]\,,
\end{equation}
for $\gamma/\beta < 0$.
Where $\gamma$ and $\beta$ are real, integration constants 
and we shall restrict our attention to the principal value of
the complex logarithm function in Eq. (\ref{10}). Based on 
\cite{brady}, we shall assume that $\Phi(u,v) \equiv 0$ for 
$v < 0$.

In terms of $r(u,v)$ Eq. (\ref{8}), and $\sigma(u,v)$ Eq. 
(\ref{9}), the line element Eq. (\ref{1}) becomes,

\begin{equation}
\label{12}
ds^2\, =\, - 2 e^{2\sigma_0} \left({r\over u}\right)^{(1-\alpha)}
du dv\, +\, r^2 d\theta^2\, .
\end{equation}
One may notice from Eqs. (\ref{8}-\ref{10}), that for different 
values of $\alpha$, $\beta$ and $\gamma$, one has different 
space-times.

Observing Eq. (\ref{12}), we notice that these space-times have a
singularity at $r=0$. It is a physical singularity as can be seen
directly from the curvature scalar $R$.

In order to show this result we start writing down the Ricci tensor
that, in the present case, has the following expression \cite{taub},

\begin{equation}
\label{13}
R_{\alpha \beta}\, =\, \Phi,_\alpha\, \Phi,_\beta \, .
\end{equation}
From it, we may compute $R$ straightforwardly with the aid of
Eqs. (\ref{8})-(\ref{12}), finding,

\begin{equation}
\label{14}
R\, =\, - 2 (1 - \alpha) \gamma \beta e^{-2\sigma_0}{[(\alpha 
v)^{1/\alpha} u]^{(1-\alpha)}\over r^{(3-\alpha)}}\, .
\end{equation}

Finally, taking the limit $r \to 0$ in $R$ Eq. (\ref{14}), we
find that this quantity diverges at $r=0$. There is no other physical
singularity for these space-times because $R$ is well defined outside
$r=0$. In particular, $u=0$ is just an apparent singularity and a new
coordinate system can be found where it disappears. 

Another important property we can learn from $R$ is the asymptotic
behavior of our solution. If we take the limit $r \to \infty$ of $R$
Eq. (\ref{14}), we find that $R \to 0$. Therefore, we conclude that 
the space-times under investigation are asymptotically flat.

Now, we would like to select the values of $\alpha$, $\beta$,
$\gamma$, such that,
our solution (\ref{8}-\ref{10}) represents the formation of
naked singularities. There are some conditions to be satisfied. 
Initially, $r$ Eq. (\ref{8}) has to be a real, positive 
function. Also, $r$ = constant 
must be a set of time-like surfaces for different constants. 
On the other hand, observing the scalar field expressions Eqs.
(\ref{9,5}) and (\ref{10}) we notice that the free parameters 
will have to be chosen in a way that $\Phi$ becomes real.

It is clear from Eqs. (\ref{12}) and (\ref{14}) that $\alpha=1$
is, locally, the three-dimensional Minkowski space-time. 
Therefore, we shall restrict our attention to space-times with
self-similarity of the second kind. Since $\Phi(u,v)\equiv 0$, for
$v < 0$, it is appropriate to consider the influx of scalar field 
to be turned on at the advanced time $v=0$. So that for $v < 0$
and $|\gamma| = 1$ the space-time is Minkowskian and the metric 
is therefore $C^1$ at the surface $v = 0$. If we let $|\gamma|
\neq 1$ the space-time for $v < 0$ would be Minkowskian only
locally. It could develop a globally non-trivial structure. For
example, if $0 < |\gamma| < 1$ the space-time for $v <0$ would
be conical with $0 < \theta < 2\pi \gamma$.

The naked singularity space-times are obtained for $\gamma = -1$,
$\beta < 0$ and $\alpha = l/m$, where $l$ and $m$ are integer
numbers. $l$ being even and $m$ odd. It means that, the scalar
field expression relevant to these space-times is Eq. (\ref{9,5}).
Figure $1$ shows a conformal diagram for a 
typical space-time in this case. We may see that the space-time
is divided, naturally, in two distinct regions. The first
one is the Minkowskian region where $v < 0$ (I). Then, we
have the collapse region where $v >0$ and $u < 0$ (II).

The scalar field Eq. (\ref{9,5}) starts collapsing from $v = 
0$, in the collapse region. From Eq. (\ref{9,5}), it is not
difficult to see that it is real and may be written as,

\begin{equation}
\label{14,5}
\Phi (u,v)\, =\, 2 (1 - \alpha)^{1/2} \tanh^{-1} 
\sqrt{{\beta(\alpha v)^{1/\alpha}\over u}}\, ,
\end{equation}
in this region. $\Phi = 0$ at $v=0$ and increases with
$v$ until it blows up at the singularity $r=0$.

In order to identify the nature of the surfaces $r=$constant 
we shall have to compute,

\begin{equation}
\label{25}
2g^{uv}\, r,_u \, r,_v\, .
\end{equation}
For the above space-times, this equation (\ref{25}) takes the
following form when we introduce the appropriate information from
Eqs. (\ref{8}) and (\ref{12}),

\begin{equation}
\label{26}
2 e^{-2\sigma_0} \beta \left[ {(\alpha v)^{1/\alpha} u
\over r}\right]^{(1-\alpha)}\, .
\end{equation}

We can see from Eq. (\ref{26}) that the surfaces $r=$constant 
will be time-like in the collapse region because with our 
choice of $l$ and $m$ the numerator of $1-\alpha=(m-l)/m$, is 
an odd integer number. Therefore, the singularity $r=0$ will 
be a time-like one. Once this singularity is not hidden by any 
horizon, we may consider these space-times representing naked 
singularities.

An important property of the naked singularities formed in 
this process comes from the fact that they are asymptotically 
flat. If we add to this the fact that they define a region of
non-zero measure in the parameter space of solutions, we can
say that the naked singularities found here, in the same way
as the ones found in \cite{kuchar}, may be considered as 
counter-examples to the weak cosmic censorship conjecture. On
the other hand, due to the fact that the $r=0$ singularity is
time-like in the present case, differently from the ones in
\cite{kuchar}, the naked singularities are also 
counter-examples to the strong cosmic censorship conjecture
\cite{penrose1}.

We may also describe our solution with the aid of the 
time coordinate,

\begin{equation}
\label{28}
t(u,v)\, =\, - \beta (\alpha v)^{1/\alpha}\, + \, \gamma u
\, ,
\end{equation}
in terms of which, the line element Eq. (\ref{12}) and the
scalar fields Eqs. (\ref{9,5}) and (\ref{10}) become, 
respectively,

\begin{equation}
\label{29}
ds^2 = -{(2)^{(1-2\alpha)}e^{2\sigma_0}\over (\beta
\gamma)^{\alpha}}\left({r\over r^2 -t^2}\right)^{(1-\alpha)}
(- dt^2 + dr^2) + r^2 d\theta^2\, ,
\end{equation}
\begin{equation}
\label{29,5}
\Phi (u,v)\, =\, 2 (\alpha - 1)^{1/2} \arctan  
\sqrt{{r-t\over r+t}}
\end{equation}
and
\begin{equation}
\label{30}
\Phi(r,t)\, =\, (1-\alpha)^{1/2}\, \ln \left[\, {\sqrt{r+t} - 
\imath \sqrt{r-t}\over \sqrt{r+t} + \imath \sqrt{r-t}}\,
\right]\, .
\end{equation}
Where we restrict our attention to the principal value 
of the complex logarithm function in Eq. (\ref{30}).

We end this article by noting that recently another
continuously self-similar solution (CSS), to the same 
problem treated here, was found in \cite{garfinkle}. 
There, it was shown to correctly describes the critical
solution, found numerically in \cite{choptuik1}, to
the collapse of a circularly symmetric, minimally coupled, 
massless, real scalar field with a negative cosmological
constant in $2+1$-dimensional general relativity. Although,
the presence of a cosmological constant prevents the 
equations to have the CSS symmetry it was shown in
\cite{choptuik1} that the critical solution has this
symmetry. Indeed, the analytical solution found in
\cite{garfinkle} has a CSS of the first kind \cite{coley}.

The critical solution separates two sets of end states
for the scalar field collapse. In the first one, the scalar
field collapses, interacts and disperses leaving behind the
anti-de Sitter space-time. In the second, black holes with 
exterior regions settling down to a BTZ form are formed.
The critical solution cannot have an apparent horizon,
therefore the singularity formed in the collapse should be
naked. The solution found in \cite{garfinkle} has no
apparent horizons, the scalar field is real and it has a null 
singularity ($u=0$ in our coordinates), in the region where 
it was compared with the numerical solution of 
\cite{choptuik1}.

It is clear from the above that our solution describes a
different physical situation than the one found in
\cite{garfinkle}.

I am grateful to A. Wang for suggestive discussions in the
course of this work. I would like also to thank
I. D. Soares for helpful discussions and FAPEMIG for the
invaluable financial support.

\begin{figure}
\psfig{file=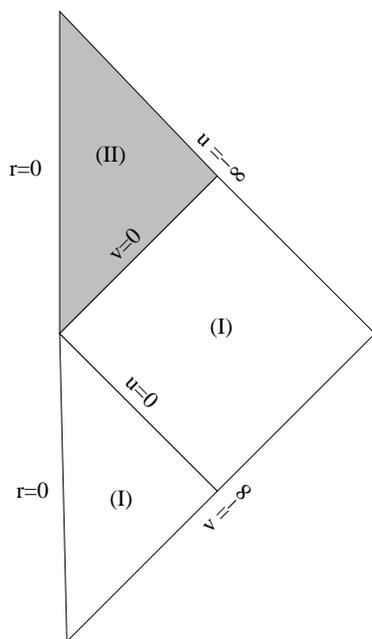}
\caption{Conformal diagram for a typical naked singularity 
solution. The null surface $v=0$ separates the
Minkowskian region (I) from the collapse region (II)  
where lies the time-like singularity at $r=0$.}
\end{figure}

\end{document}